\documentstyle{aipproc}
\begin{document}
\title{Muon-proton Colliders: Leptoquarks and Contact Interactions
\footnote{Invited talk present at the Fourth International Conference on
the Physics Potential and Development of $\mu^+ \mu^-$ Colliders, 
San Francisco CA, Decmember 1997.}
}
\author{Kingman Cheung}
\address{
Department of Physics, University of California at Davis, 
Davis, CA 95616}
\maketitle

\begin{abstract}
The muon-proton ($\mu p$) collider is an interesting option of the muon 
collider.   Here we discuss the physics potential of the 
$\mu p$ collider; especially, leptoquarks and contact interactions.  We 
calculate the sensitivity reach for the second generation leptoquarks and 
leptogluons, $R$-parity violating squarks, and $\mu q$ contact  interactions
for the $\mu p$ colliders of various energies and luminosities.
\end{abstract}

\section*{Introduction}

Recently, the muon collider has 
received a lot of attentions \cite{rd}.  The muon collider of
a few hundred GeV center-of-mass energy is considered a Higgs factory
\cite{report},
where interactions and branching ratios of the Higgs boson can be studied
in detail.  It is also an excellent place to study the
top-quark near the threshold region \cite{precision}.  
Other physics opportunities include
precision studies of gauge bosons \cite{precision}, 
search for supersymmetry and 
lepton-number violation, and other new physics.  Muon colliders in TeV
range should be feasible for studying strong electroweak symmetry breaking
\cite{strong},
lepton-number violation, and search for heavy exotic particles.

The R\&D \cite{rd,fmc} of the muon collider is underway.  The First Muon 
Collier (FMC) will have a 200 GeV muon beam
 on a 200 GeV anti-muon beam, which could
possibly be at the Fermilab \cite{fmc}.
With the existing Tevatron proton beam the
muon-proton collision becomes a possible option.  It would be a 200 GeV
$\otimes$ 1 TeV $\mu p$ collider, where the first energy is the energy
of the muon beam and the second energy is the proton beam energy.  The
existing lepton-proton collider is the $ep$ collider at HERA.
Lepton-proton colliders have been proved to be successful by the physics
results from HERA.  In this work, we shall discuss the physics
potential of the $\mu p$ colliders at various energies and
luminosities.  Other $\mu p$ colliders that we are considering are 50
GeV $\otimes$ 1 TeV, 1 TeV $\otimes$ 1 TeV, and 2 TeV $\otimes$ 3 TeV.
The center-of-mass energies and luminosities of these various
designs are summarized in Table \ref{table1}.  The nominal yearly
luminosity of the 200 GeV $\otimes$ 1 TeV $\mu p$ collider is
about 13 fb$^{-1}$.  Luminosities for other designs are roughly scaled
by one quarter power of the muon beam energy and given in Table
\ref{table1}.

\begin{table}[b!]
\caption{
The center-of-mass energies $\sqrt{s}$ and luminosities ${\cal L}$ 
of various designs of muon-proton colliders.}
\label{table1}
\begin{tabular*}{5.8in}{@{\extracolsep{1in}}ccc}
& $\sqrt{s}({\rm GeV})$  & ${\cal L}\;({\rm fb}^{-1})$ \\
\tableline
\tableline
$30{\rm GeV} \otimes 820{\rm GeV}$ &  314 & 0.1 \\
$50{\rm GeV} \otimes 1{\rm TeV}$ &    447 & 2 \\
$200{\rm GeV} \otimes 1{\rm TeV}$  &  894 & 13 \\
$1{\rm TeV} \otimes 1{\rm TeV}$ &     2000 & 110 \\
$2{\rm TeV} \otimes 3{\rm TeV}$  &    4899 & 280 \\
\end{tabular*}
\end{table}

\section*{Physics Potential}

The physics opportunities of $\mu p$ colliders are similar to
those of $ep$ colliders, but the sensitivity reach might be very different,
which depends on how precise the particles can be identified and measured
in $ep$ and $\mu p$ environments.  
Similar to $ep$ colliders the proton structure functions can be measured 
to very large $Q^2$ and small $x$ in $\mu p$ colliders of 
higher energies.  At HERA, the $Q^2$ has been measured up to $Q^2 \simeq
10^4$ GeV$^2$ and $x$ down to $x \sim 3\times 10^{-4}$.  At the 
200 GeV $\otimes$ 1 TeV $\mu p$ collider the $Q^2$ can be measured up to 
$10^6$ GeV$^2$. In addition,  QCD studies, 
search for supersymmetry and other exotic 
particles can be carried out at the $\mu p$ colliders.  Here we are 
particularly interested in the leptoquarks, leptogluons, $R$-parity 
violating squarks, and the contact interactions that are specific to
the muon.  The goal here is to estimate the sensitivity reach for these
new physics at various energies and luminosities.

\subsection*{Leptoquarks}

The second generation leptoquarks made up of a muon and 
a charm or strange quark are particularly interesting at the $\mu p$
collider because they 
can be directly produced in the $s$-channel processes, e.g.,
\begin{equation}
\mu^\pm  c \to L^0_{\mu c} \;.
\end{equation}
It is conventional to assume no inter-generational mixing in order to 
prevent the dangerous flavor-changing neutral currents.  The $s$-channel
production will give spectacular enhancement in the invariant mass $M$ of
the muon and the hadronic final state, or the $x=s/M^2$ distribution.

The Lagrangian of the second generation leptoquark with the muon and charm 
and strange quarks is given by
\begin{equation}
{\cal L} = \lambda^0_{\mu q} \bar q \mu L^0_{\mu q} +
           \lambda^1_{\mu q} \bar q \gamma_\rho \mu L^{1\rho}_{\mu q} +
           {\rm h.c.} \;,
\end{equation}
where $q=c,s$ and the superscripts $(0,1)$ on the leptoquark field denote
the scalar
and the vector leptoquarks, respectively.  The production cross section
of the leptoquark at the $\mu p$ collider is given by
\begin{equation}
\sigma = \frac{\pi \lambda^2}{4 s} \; q(x,Q^2) \times (J+1) \;,
\end{equation}
where $J$ is the spin of the leptoquark and $q(x,Q^2)$ is the parton
luminosity.

\subsection*{Leptogluons}

A leptogluon has a spin of either $1/2$ or $3/2$, a lepton quantum number
(in this case it is the muon), and a color quantum number (the same as gluon.)
The interaction Lagrangian for a spin $1/2$ leptogluon is given by
\begin{equation}
{\cal L} = g_s \frac{M_{L_{\mu g}}}{2 \Lambda^2} \overline{ L^a_{\mu g} }
\sigma^{\mu \nu} \mu \, G_{\mu \nu}^b \, \delta_{a b}  + {\rm h.c.} \;,
\end{equation} 
where $\Lambda$ is the scale that determines the strength of the interaction.
The decay width of the leptogluon into a muon and a gluon is given by
\begin{equation}
\Gamma(L_{\mu g}\to \mu g) = \frac{\alpha_s M^5_{L_{\mu g}} }{2 \Lambda^4} \;.
\end{equation}
The leptogluon can be produced in $s$-channel in a $\mu p$ collider and 
the production cross section is given by
\begin{equation}
\sigma = \frac{4 \pi^2 \alpha_s}{s}\,
    \left( \frac{M^2_{L_{\mu g}}}{\Lambda^2} \right )^2
    g(x,Q^2) \;,
\end{equation}
where $g(x,Q^2)$ is the gluon luminosity.

\subsection*{$R$-parity Violating Squarks}
$R$-parity is in general assumed in supersymmetry, but there is no
theoretical reasons why $R$-parity should conserve.  
$R$-parity violation is included by introducing additional terms in
the superpotential:
\begin{equation}
{\cal W}_{\not{R}} = \lambda_{ijk} L_i L_j \overline{E_k}
 + \lambda'_{ijk} L_i Q_j \overline{D_k} 
 + \lambda^{''}_{ijk} \overline{U_i}\, \overline{D_j}\, \overline{D_k} 
 + \mu_i L_i H_u \;,
\end{equation}
where $L,\overline{E}, Q, \overline{U},\overline{D}, H_u$ are superfields.
The relevant term in the superpotential for the direct production of the 
$R$-parity violating squark at the $\mu p$ collider is $\lambda'_{ijk} L_i Q_j
\overline{D_k}$.  The corresponding Lagrangian is 
\begin{eqnarray}
{\cal L}_{ L_i Q_j \overline{D_k}} & =& \lambda'_{ijk} \biggr[
\tilde{e}_{iL} \overline{d_{kR}} u_{jL} +
\tilde{u}_{jL} \overline{d_{kR}} e_{iL} +
\tilde{d^*}_{kR} \overline{(e_{iL})^c} u_{jL} \nonumber \\
& & -
\tilde{\nu}_{iL} \overline{d_{kR}} d_{jL}
-\tilde{d}_{jL} \overline{d_{kR}} \nu_{iL}
-\tilde{d^*}_{kR} \overline{(\nu_{iL})^c} d_{jL} \biggr] + h.c. \;
\label{eq:Lyukawa}
\end{eqnarray}
where $i,j,k$ are the family indices, and $c$ denotes the charge conjugate.
The $R$-parity violating squarks can be considered special scalar leptoquarks.

The cross section for $\mu^+ p \to \tilde{t}_L \to \mu^+ X$ is given by
\begin{equation}
\label{90}
\sigma_{\tilde{t}_L} = \frac{\pi |\lambda'_{231}|^2}{4s} \; d \left(
\frac{m^2_{\tilde{t}_L}}{s}, Q^2=m_{\tilde{t}_L}^2 \right ) \;,
\end{equation}
where $d$ is the down-quark luminosity.  
The above formula can be easily changed to represent the production of
other squarks with the corresponding subscripts in $\lambda'$ and the
parton luminosity. 
If kinematically allowed the leptoquarks, leptogluons, and the $R$-parity
violating squarks are produced in $s$-channel and thus give rise to 
spectacular enhancement in a single bin of the invariant mass $M$ distribution
or the $x$ distribution.

\subsection*{Contact Interactions}
The effective four-fermion contact interactions can arise from 
fermion compositeness or exchanges of heavy particles like heavy $Z'$,
heavy leptoquarks, or other exotic particles.
The conventional effective Lagrangian of $ll q q$ ($l=e,\mu$) contact
interactions has the form \cite{me}
\begin{eqnarray}
L_{NC} &=& \sum_q \Bigl[ \eta_{LL}
\left(\overline{l_L} \gamma_\mu l_L\right)
\left(\overline{q_L} \gamma^\mu q_L \right)
+ \eta_{RR} \left(\overline{l_R}\gamma_\mu l_R\right)
                 \left(\overline{q_R}\gamma^\mu q_R\right) \nonumber\\
&& \qquad {}+ \eta_{LR} \left(\overline{l_L} \gamma_\mu l_L\right)
                             \left(\overline{q_R}\gamma^\mu q_R\right)
+ \eta_{RL} \left(\overline{l_R} \gamma_\mu l_R\right)
\left(\overline{q_L} \gamma^\mu q_L \right) \Bigr] \,, \label{effL}
\end{eqnarray}
where the eight independent coefficients $\eta_{\alpha\beta}^{lu}$ and
$\eta_{\alpha\beta}^{ld}$ have dimension (TeV)$^{-2}$ and are conventionally
expressed as $\eta_{\alpha\beta}^{lq} = \epsilon g^2 /\Lambda_{lq}^2$,
with a fixed $g^2=4\pi$.

We introduce the reduced amplitudes 
$M^{\mu q}_{\alpha \beta}$, where the subscripts label the chiralities
of the initial lepton ($\alpha$) and quark ($\beta$). The SM tree level reduced
amplitude for $\mu q\to \mu q$ is
\begin{equation}
\label{reduced}
M^{\mu q}_{\alpha\beta}(\hat t) = -\frac{e^2 Q_q}{\hat t} +
\frac{e^2 }{\sin^2 \theta_{\rm w}  \cos^2 \theta_{\rm w}}
\frac{g_\alpha^\mu g_\beta^q}{\hat t - m_Z^2} \;,
\qquad \alpha,\beta = L,R
\end{equation}
where $\hat t = -Q^2$ is the Mandelstam variable,
$g_L^f = T_{3f} - \sin^2 \theta_{\rm w} Q_f$ and
$g_R^f = -\sin^2 \theta_{\rm w} Q_f$, $T_{3f}$ and $Q_f$ are, respectively,
the third component of the SU(2) isospin and the electric charge of the
fermion $f$ in units of the proton charge, and $e^2=4\pi \alpha_{\rm em}$.
The new physics contributions to the reduced amplitudes $M_{\alpha\beta}$ from
the $\mu \mu qq$ contact interactions of Eq.~(\ref{effL}) are
\begin{equation}
\Delta M^{\mu q}_{\alpha \beta}  = \eta^{\mu q}_{\alpha \beta}\;, \qquad
\alpha,\beta = L,R \;. \label{newphys}
\end{equation}
The differential cross section are given by
\begin{eqnarray}
{d\sigma(\mu^+p)\over dx\,dy} &=& {sx\over16\pi} \left\{ u(x,Q^2) \left[
\left| M_{LR}^{\mu u} \right|^2 + \left| M_{RL}^{\mu u} \right|^2 + (1-y)^2 \left(
\left| M_{LL}^{\mu u} \right|^2 + \left| M_{RR}^{\mu u} \right|^2 \right) \right]
\right. \nonumber\\
&&\quad\left. {}+ d(x,Q^2) \left[ \left| M_{LR}^{\mu d} \right|^2 + \left|
M_{RL}^{\mu d} \right|^2 + (1-y)^2 \left( \left| M_{LL}^{\mu d} \right|^2 + \left|
M_{RR}^{\mu d} \right|^2 \right) \right] \right\}  \label{bar-ep}\\
{d\sigma(\mu^-p)\over dx\,dy} &=& {sx\over16\pi} \left\{ u(x,Q^2) \left[ \left|
M_{LL}^{\mu u} \right|^2 + \left| M_{RR}^{\mu u} \right|^2 + (1-y)^2 \left( \left|
M_{LR}^{\mu u} \right|^2 + \left| M_{RL}^{\mu u} \right|^2 \right) \right] \right.
\nonumber\\
&&\quad\left. {}+ d(x,Q^2) \left[ \left| M_{LL}^{\mu d} \right|^2 + \left|
M_{RR}^{\mu d} \right|^2 + (1-y)^2 \left( \left| M_{LR}^{\mu d} \right|^2 + \left|
M_{RL}^{\mu d} \right|^2 \right) \right] \right\}
\end{eqnarray}
The above contact interactions do not enhance the cross section in a single
bin of the invariant mass distribution like the light leptoquarks do, instead,
contact interactions enhance the cross section at the large $Q^2$ tail. 

\section*{Sensitivity Reach}

\begin{table}[b!]
\caption{\label{table2}
The 95\% sensitivity of the $\Lambda_{\alpha\beta}^{\mu q},\;
(\alpha,\beta=L,R ;\, q=u,d)$ that can be reached at the various $\mu^+ p$ 
colliders, by assuming that what will be observed is given by the SM 
prediction.}
\begin{tabular}{c@{\extracolsep{-0.4in}}||cc|cc|cc|cc|cc}
&
\multicolumn{2}{|c@{\extracolsep{-0.1in}}|}{$30{\rm GeV} \otimes 820{\rm GeV}$} &
\multicolumn{2}{|c@{\extracolsep{-0.1in}}|}{$50{\rm GeV} \otimes 1{\rm TeV}$} &
\multicolumn{2}{|c@{\extracolsep{-0.1in}}|}{$200{\rm GeV} \otimes 1{\rm TeV}$}  &
\multicolumn{2}{|c@{\extracolsep{-0.1in}}|}{$1{\rm TeV} \otimes 1{\rm TeV}$} &   
\multicolumn{2}{|c@{\extracolsep{-0.1in}}}{$2{\rm TeV} \otimes 3{\rm TeV}$}  \\
\tableline
$\sqrt{s}({\rm GeV})$ & \multicolumn{2}{|c|}{314} & 
                        \multicolumn{2}{|c|}{447} & 
                        \multicolumn{2}{|c|}{894} & 
                        \multicolumn{2}{|c|}{2000} & 
                        \multicolumn{2}{|c}{4899} \\
${\cal L}\;({\rm fb}^{-1})$ &\multicolumn{2}{|c|}{0.1} & 
                            \multicolumn{2}{|c|}{2}   &
                            \multicolumn{2}{|c|}{13}   &
                            \multicolumn{2}{|c|}{110}   &
                            \multicolumn{2}{|c}{280}    \\
\tableline
\tableline
 & $+$ & $-$ &  $+$ & $-$ &  $+$ & $-$ &  $+$ & $-$ & $+$ & $-$\\
$\Lambda_{LL}^{\mu u}$ & 
1.4 & 0.8     &  3.9 & 3.7     & 9.4 & 9.2   & 24 & 23    & 46 & 46 \\
$\Lambda_{LR}^{\mu u}$ & 
2.0 & 1.3     &  4.9 & 4.3     & 11 & 9.2   & 26 & 23    &  50 & 40  \\
$\Lambda_{RL}^{\mu u}$ & 
1.9 & 1.3     &  4.3 & 3.2     & 8.9 & 6.4   & 21 & 14   &  40 & 29  \\
$\Lambda_{RR}^{\mu u}$ & 
1.4 & 0.8     &  3.5 & 3.2     & 8.4 & 7.9   & 21 & 20    & 40 & 37 \\
\tableline
$\Lambda_{LL}^{\mu d}$ & 
0.7 & 1.1     &  2.2 & 2.7     & 6.2 & 6.6   & 17 & 17    &  32 & 34  \\
$\Lambda_{LR}^{\mu d}$ & 
1.1 & 1.3     &  2.1 & 2.8     & 4.5 & 6.0   & 10 & 14    & 31  & 29  \\
$\Lambda_{RL}^{\mu d}$ & 
1.2 & 1.2     &  2.5 & 2.2     & 5.6 & 4.4   & 14 & 10    & 29  & 22  \\
$\Lambda_{RR}^{\mu d}$ & 
0.8 & 1.0     &  1.5 & 2.4     & 3.7 & 5.1   & 9.9 & 13    & 18 & 25 \\
\end{tabular}
\end{table}

The 95\% sensitivity of the contact interaction scale that can be reached by
$\mu p$ colliders at various center-of-mass energies and luminosities are 
performed in the following.  We use the $x$-$y$ distribution to investigate
the sensitivity to the new contact interactions.
We divide the $x$-$y$ plane into a grid with $0.05<x<0.95$ and $0.05<y<0.95$
and 0.1 interval in both $x$ and $y$ directions.  We calculate the number of 
events predicted by the standard model in each bin, call it $n_i^{\rm sm}$.  
We use an overall efficiency of 0.8.
We assume that the observed number of events is given by the standard model.  
We vary one $\eta_{\alpha\beta}^{\mu q}$ at a time while 
keeping others zero and calculate the 
predicted number of events in each bin, call it $n_i^{\rm th}$.  
We then calculate the $\chi^2$ using 
\begin{equation}
\chi^2 = \sum_{i} \biggr[ 2 ( n_i^{\rm th} - n_i^{\rm sm}) + 
2 n_i^{\rm sm} \log \left(
\frac{n_i^{\rm sm}}{n_i^{\rm th}} \right ) \biggr ] \;,
\end{equation}
where the sum is over all $9\times 9$ bins.
We know that for a larger $\eta$ we will obtain a larger $\chi^2$, which means
that it is really different from the standard model beyond statistical 
fluctuation. 
Here we have 80 degree of freedom, and so for a 95\% CL we set $\chi^2=102$.
We then repeat for another $\eta$.

The sensitivity reach of $\Lambda^{\mu q}_{\alpha \beta}$ is tabulated in
Table \ref{table2}.  The sensitivity reach depends on the sign of the 
contact term.  The maximum reach of $\Lambda$ at each center-of-mass 
energy  roughly scales as $\Lambda \sim 10 \sqrt{s}$.
The effect of luminosity on $\Lambda$ is rather small: $\Lambda$ only 
scales as the $1/4$th power of the luminosity.

\begin{table}[b!]
\caption{\label{table3}
95\% sensitivity on $\lambda'_{231}$ for a few choices of 
$m_{\tilde{t}_L}$ at various $\mu^+ p$ colliders.}
\begin{tabular}{c@{\extracolsep{-0.2in}}||c@{\extracolsep{-0.1in}}|c
@{\extracolsep{-0.1in}}|c@{\extracolsep{-0.1in}}|c@{\extracolsep{-0.1in}}|c
@{\extracolsep{-0.1in}}}
&$30{\rm GeV} \otimes 820{\rm GeV}$ & $50{\rm GeV} \otimes 1{\rm TeV} $
&$200{\rm GeV} \otimes 1{\rm TeV}$  & $1{\rm TeV} \otimes 1{\rm TeV} $
&$2{\rm TeV} \otimes 3{\rm TeV}$ \\
\tableline
$\sqrt{s} ({\rm GeV})$ & 314 & 447  & 894 & 2000 & 4899 \\
\hline
${\cal L} ({\rm fb}^{-1})$ & 0.1 & 2 & 13 & 110 & 280 \\
\tableline 
\tableline
$m_{\tilde{t}_L} \; ({\rm GeV})$ & & & & & \\
200 & 0.014 & 0.0045 & 0.0025& 0.0015 & 0.0010 \\ 
300 &  -    & 0.0094 & 0.0032& 0.0018 & 0.0014 \\
400 &  -    & 0.055  & 0.0041& 0.0021 & 0.0017 \\
500 &  -    &   -    & 0.0056& 0.0024 & 0.0019 \\
600 &  -    &   -    & 0.0086& 0.0027 & 0.0021 \\
700 &  -    &   -    & 0.016 & 0.0030 & 0.0023 \\
800 &  -    &   -    & 0.045 & 0.0033 & 0.0025  \\
900 &  -    &   -    &   -   & 0.0037 & 0.0026  \\
1000 & -    &   -    &   -   & 0.0043 & 0.0027  \\
1500 & -    &   -    &   -   & 0.012  & 0.0034  \\
2000 & -    &   -    &   -   &   -    & 0.0043  \\
2500 & -    &   -    &   -   &   -    & 0.0056  \\
3000 & -    &   -    &   -   &   -    & 0.0078 \\
3500 & -    &   -    &   -   &   -    & 0.013 \\
4000 & -    &   -    &   -   &   -    & 0.024 \\
4500 & -    &   -    &   -   &   -    & 0.081 \\
\end{tabular}
\end{table}

To estimate the sensitivity reach for $R$-parity violating squarks 
we start with $\lambda'_{231}$ for the top squark and the down quark 
luminosity.
We assume the enhancement in cross section is in 
the mass bin of $0.9\; m_{\tilde{t}_L} <
m < 1.1\; m_{\tilde{t}_L}$.  We calculate the number of events predicted by
the standard model in this bin, call it $n^{\rm sm}$.  Again, we use an 
overall efficiency of 0.8.  Then we use the poisson statistics to
estimate the $n^{\rm th}$ that $n^{\rm sm}$ can fluctuate to at the 95\% CL:
\begin{equation}
\sum_{n=0}^{n^{\rm th}} \frac{(n^{\rm sm})^n e^{-n^{\rm sm}} } { n !}  > 0.95
\end{equation}
where $n^{\rm th}$ is the first $n$ that the above inequality is satisfied. 
Once the $n^{\rm th}$ is obtained the $\lambda'_{231}$ can be obtained using
Eq. (\ref{90}).

The sensitivity reach of $\lambda_{231}'$ is tabulated in Table ~\ref{table3}.
We have also calculated the sensitivity reach of $\lambda_{232}'$ using the
strange quark luminosity.  We found that the reach is typically worse than
that of $\lambda_{231}'$: for small $m_{\tilde{t}_L}$ the reach is about a
factor of two worse while for large $m_{\tilde{t}_L}$ the reach can be 
ten times worse.  This is because the strange quark luminosity is rather 
large at small $x$ but very small at large $x$.


The results for the second generation leptoquarks and leptogluons are
summarized in Table \ref{table4} \cite{new}.  
Here only a simple criteria is defined
for the discovery of the leptoquarks and leptogluons.  
Assuming no background and requiring five signal events 
for the discovery the sensitivity reach is at 99\% CL. 

\begin{table}[b!]
\caption{\label{table4}
99\% sensitivity reach on the coupling $\lambda$ for the second generation
leptoquarks and the new physics scale $\Lambda$ for leptogluon via the
resonance production $\mu p \to L$.}
\begin{tabular}{c||c|c|c|c}
& $30{\rm GeV} \otimes 820{\rm GeV}$ & $50{\rm GeV} \otimes 1{\rm TeV}$ 
& $200{\rm GeV} \otimes 1{\rm TeV}$  & $2{\rm TeV} \otimes 4{\rm TeV} $\\
\tableline
$\sqrt{s} ({\rm GeV})$ & 314 & 447  & 894 & 5657 \\
\tableline
${\cal L} ({\rm fb}^{-1})$ & 0.1 & 2 & 13 & 280 \\
\tableline 
\tableline
$M^0_{\mu c}$ &      200 GeV & 300 GeV & 500 GeV & 1500 GeV\\
$\lambda^0_{\mu c}$ &   0.089     & 0.043         &  0.010        & 0.0014 \\
\tableline
$M^0_{\mu s}$ &      200 GeV & 300 GeV & 500 GeV & 1500 GeV\\
$\lambda^0_{\mu s}$ &   0.068     & 0.034         &  0.0080       & 0.0011 \\
\tableline
\tableline
$M^1_{\mu c}$ &      200 GeV & 300 GeV & 500 GeV & 1500 GeV\\
$\lambda^1_{\mu c}$ &   0.063     & 0.031         &  0.0072       & 0.0010 \\
\tableline
$M^1_{\mu s}$ &      200 GeV & 300 GeV & 500 GeV & 1500 GeV\\
$\lambda^1_{\mu s}$ &   0.048     & 0.024         &  0.0055       & 0.0008 \\
\tableline
\tableline
$M_{\mu g}$   &  200 GeV & 300 GeV  & 500 GeV  & 1500 GeV\\
$\Lambda_{\mu g} ({\rm TeV})$ & 20  & 49 & 190 & 1700 \\
\tableline
\end{tabular}
\end{table}

\end{document}